\documentclass[superscriptaddress, twocolumn,showpacs,preprintnumbers,amsmath,amssymb,prl]{revtex4}

\usepackage{graphicx}
\usepackage{dcolumn}
\usepackage{bm}

\newcommand{\gammadot}{\ensuremath{\dot{\gamma}}}

\begin{document}


\title{From Frictional to Viscous Behavior: Three Dimensional Imaging and
Rheology of Gravitational Suspensions}


\author{Joshua A. Dijksman}

\affiliation{Kamerlingh Onnes Lab, Universiteit Leiden, Postbus
9504, 2300 RA Leiden, The Netherlands}

\author{Elie Wandersman}

\affiliation{Kamerlingh Onnes Lab, Universiteit Leiden, Postbus
9504, 2300 RA Leiden, The Netherlands}

\author{Steven Slotterback}

\affiliation{Department of Physics, IPST, and IREAP, University of
Maryland, College Park, Maryland 20742, USA}

\author{Christian R. Berardi}

\affiliation{Department of Physics, IPST, and IREAP, University of
Maryland, College Park, Maryland 20742, USA}

\author{William Derek Updegraff}

\affiliation{Department of Physics, IPST, and IREAP, University of
Maryland, College Park, Maryland 20742, USA}

\author{Martin van Hecke}

\affiliation{Kamerlingh Onnes Lab, Universiteit Leiden, Postbus
9504, 2300 RA Leiden, The Netherlands}

\author{Wolfgang Losert}
\affiliation{Department of Physics, IPST, and IREAP, University of
Maryland, College Park, Maryland 20742, USA}

\date{\today}

\begin{abstract}


We probe the three dimensional flow structure and rheology of
gravitational (non-density matched) suspensions for a range of
driving rates in a split-bottom geometry. We establish that for
sufficiently slow flows, the suspension flows as if it were a dry
granular medium, and confirm recent theoretical modelling on the
rheology of split-bottom flows. For faster driving, the flow
behavior is shown to be consistent with the rheological behavior
predicted by the recently developed ``inertial number'' approaches
for suspension flows.

\end{abstract}

 \maketitle

Flows of granular materials submersed in a liquid of unequal
density have started to attract considerable
attention~\cite{1999_cracadparis_ancey,
2009_rheolacta_lemaitre,2005_physfluids_cassar,2006_pre_siavoshi,2009_jfm_pailha}
and are relevant in many practical
applications~\cite{2009_science_frey}. These materials, which we
will refer to as ``gravitational'' suspensions, clearly differ
from density matched suspensions, which have been studied in great
detail~\cite{1991_powtechn_jomha,
2005_annurevfluidmech_stickel,2009_prl_fall,1906_annphys_einstein}.
Gravitational suspensions exhibit sedimentation, large packing
fractions and jamming of the material, which suggests a
description similar to dry granular
matter~\cite{1996_revmodphys_jaeger,2008_annurevfluid_forterre}.

In the last two decades, various flow regimes have been identified
for dry granular matter. Sufficiently slow flows are {\em
frictional}: the ratio of shear (driving) to normal (confining)
stresses becomes independent of flow rate if the material is
allowed to dilate~\cite{1998_physfluids_tardos,
2008_annurevfluid_forterre}. Faster flows are referred to as {\em
inertial}: here the effective friction coefficient $\mu$ depends
on the so-called ''inertial'' number $I$, which is a
non-dimensional measure of the local flow
rate~\cite{2004_epje_gdrmidi,2008_annurevfluid_forterre,2006_nature_jop}.

For gravitational suspensions, the presence of liquid instead of
gas as interstitial medium strongly affects the microscopic
picture --- how should we think of the flow of such suspensions?
Pouliquen and coworkers proposed that the ratio of the strain rate
and settling time, $I_S$, would play a similar role as the
inertial number in dry granular flows~\cite{2009_jfm_pailha}. They
furthermore conjectured a dependence of the effective friction
coefficient $\mu$ on $I_S$ similar to the dry case, and applied
this rheological law to capture the behavior of underwater
avalanches~\cite{2008_physfluid_pailha}.

Here we test this picture by combining 3D imaging and rheological
measurements of the flow of gravitational suspensions in a
so-called split-bottom geometry (Fig.~1). This geometry has two
main advantages. First, the flow rate, which is the key control
parameter in the inertial number framework, can be varied over
several orders of magnitude, allowing us to access slow flows as
seen in plane shear~\cite{2006_pre_siavoshi,2007_prl_divoux},
faster flows as seen in gravity driven
flows~\cite{2009_jfm_pailha,2007_jfm_doppler}, and the crossover
regime in between - something not achieved in previous studies of
gravitational suspensions
\cite{2006_pre_siavoshi,2007_prl_divoux,2009_jfm_pailha,2007_jfm_doppler}.
Second, extensive experimental and numerical
work~\cite{2003_nature_fenistein,2004_prl_fenistein,2006_prl_fenistein,2006_prl_cheng,
2007_epl_depken, 2007_pre_ries} has shown that the split bottom
geometry produces highly nontrivial slow dry granular flows. A
simple frictional picture is not sufficient to capture these flows
\cite{2010_softmatter_dijksman,2008_epl_sakaie}, so that testing
whether these profiles also arise in slowly sheared gravitational
suspensions is a stringent test for similarities between slow dry
flows and slow gravitational suspension flows.

\begin{figure}[t!]
    \begin{center}
        \includegraphics[width=8cm]{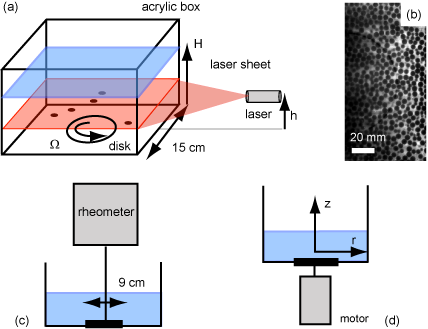}
        \caption{(Color online) (a) The experimental setup used for flow visualization in a "`split-bottom"' geometry.
          (b) Example image of a single cross section, displaying half the box. (c)
          Geometry used for rheological measurements.(d) Geometry used for flow visualization.
          }
        \label{fig:setup}
    \end{center}
\end{figure}

{\em Setup ---} The split-bottom geometry is sketched in
Fig.~\ref{fig:setup}a, and consists of a square box, 15 cm in
width with transparent acrylic walls, at the (rough) bottom of
which a (rough) disk of radius $R_s=4.5$ cm can be rotated at rate
$\Omega$.

We use monodisperse acrylic spheres with a diameter $d$ of 4.6 mm
(Engineering Labs); all our results are qualitatively similar for
3.2 mm particles. The particles are suspended in a mixture of some
78\% Triton X-100, 13 \% water, and 9\% ZnCl2 (by
weight)~\cite{2011_revsciinstr_dijksman} with a fluorescent dye
added (Nile Blue 690). The refractive indices of particles and
fluid are approximately 1.49 and match closely
--- we adapted the recipe from~\cite{1996_jfm_krishnan}. The fluids
viscosity $\eta_f$ is 0.3 ($\pm 0.05$) Pa~s, and the difference in
density between the fluid and the particles is about 100 kg/m$^3$.

The particle motion is visualized by illuminating the suspension
with a thin ( $< 200 \mu$m ) laser sheet~\cite{2006_pre_siavoshi,
2003_prl_tsai, 2008_prl_slotterback}. The laser (Stocker Yale, 635
nm) is aligned parallel to the bottom of the box
(Fig.~\ref{fig:setup}a) and mounted on a z-stage which allows the
illumination of slices of the suspension at different heights
$h$~\cite{2008_prl_slotterback}. Image acquisition is done with a
triggered 12 bit cooled CCD camera, and contrast is sufficient to
image half of the box (Fig.~\ref{fig:setup}b). We use a Particle
Image Velocimetry-like (PIV) method to obtain the normalized
azimuthal velocity $\omega(r,z) = v_{\theta}/(r\Omega)$ in slices
of constant $z$. Combining these slices, we reconstruct the full
angular velocity field as function of radius and depth for a range
of driving rates. An overview of the imaging technique will be
published elsewhere~\cite{2011_revsciinstr_dijksman}.

Rheological experiments were carried out by driving the disk from
above with a rheometer (Anton Paar MCR 501) --- see
Fig.~\ref{fig:setup}c. Velocimetry measurements were done by
driving the disk from below with a DC motor --- see
Fig.~\ref{fig:setup}d. There is always at least half a centimeter
of fluid above the suspension to ensure that the surface tension
of the fluid will not affect the
dilation~\cite{1998_physfluids_tardos} of the packing.

{\em Constitutive equation ---} We derive the constitutive
equation for our suspension in the modified ''inertial number''
approach~\cite{2005_physfluids_cassar}. The typical rearrangement
timescale for the particles in the suspension, given the viscosity
and relative density of the particles becomes $d/v_{\inf} =
\eta_f/P\alpha$, where $v_{\inf}$, $P$ and $\alpha$ are settling
velocity, pressure and porosity, so that the 'inertial number'
becomes: $I_S = \frac{\dot{\gamma}\eta_f}{P\alpha}$
~\cite{2005_physfluids_cassar}. The shear stress stress $\tau$ is
then written as $\tau = \mu(I)P$, with $\mu(I)$ an empirical
friction function. For small $I$, $\mu(I)$ can be expanded:
$\mu(I) = \mu_0 + \mu_1 I$
~\cite{2005_physfluids_cassar,2009_pre_koval, 2009_jfm_pailha},
with $\mu_0$ and $\mu_1$ empirical values. Combining this with the
expression for $I_S$, we arrive at:
\begin{equation}
\tau = \mu_0P + \mu_1\frac{\eta_f\dot{\gamma}}{\alpha}
\label{eq:tau}
\end{equation}
\noindent  Thus, to lowest order, the local stress in a
gravitational suspension is a linear combination of a frictional
stress and a purely viscous stress~\cite{2009_jfm_pailha}. This is
reminiscent of the rheology of a Bingham fluid, in that slow flows
are rate independent while faster flows become dominated by simple
viscous drag. There is, however, a crucial difference: for slow
driving, the shear stresses are predicted to be {\em proportional}
to the pressure, while only for faster flows, the shear stresses
become asymptotically {\em independent} of pressure.

\begin{figure}[t!]
    \includegraphics[width= 8.5cm]{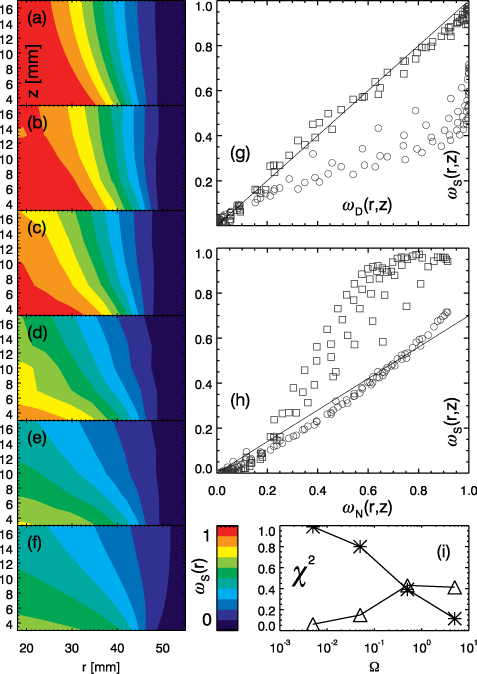}
    \caption{(Color online) Frictional to viscous crossover in flow profiles. (a) The predicted flow field
            for dry granular flows $\omega_D(r,z)$ from Refs.~\cite{2003_nature_fenistein,2004_prl_fenistein,
            2006_prl_fenistein,2006_prl_cheng,2007_epl_depken, 2007_pre_ries,2010_softmatter_dijksman}.
            (b-e) Measured velocity fields $\omega_S(r,z)$ at driving rates $\Omega = 8.3\times 10^{-5}$ rps (b),
            $ \Omega = 8.3\times10^{-4}$ rps (c) , $8.3\times10^{-3}$ rps (d)
            and $8.3\times10^{-2}$ rps (e). (f) The Newtonian flow field
            $\omega_N(r,z)$ calculated with the finite element
            method (see text). Note the similarity of (a) to (b)
            and (e) to (f). (g) Scatter plot comparison of $\omega_D(r,z)$ (a)
            and $\omega_S(r,z)$ for $\Omega = 8.3\times 10^{-5}$ rps
            ($\Box$)(a) and $\Omega = 8.3\times 10^{-2}$ rps ($\circ$).
            (h) Scatter plot comparison of $\omega_S(r,z)$ and the flow field of a Newtonian flow
            $\omega_N(r,z)$ for  $\Omega = 8.3\times 10^{-5}$ rps ($\Box$)
            and $\Omega = 8.3\times 10^{-2}$ rps ($\circ$).
            (i) $\chi^2$ vs $\Omega$ for comparison to granular ($\triangle$)
            and Newtonian flow ($\star$).  The flow characteristics change
            from granular to Newtonian with increasing shear rate.
            \label{fig:submstruct:concoverview}}
\end{figure}

{\em Flow profiles ---} 
%
In Fig.~\ref{fig:submstruct:concoverview}a-f we compare, for a
range of driving rates, the measured flow fields, $\omega_S(r,z)$
(panel b-e) with predicted flow fields for dry granular media
(panel a) and Newtonian flow (panel f). We fix the particle
filling height at 23 mm $(H/R_S \approx 0.5)$. Clearly the flow
structure progressively changes for faster flow rates, as the
defining characteristic of slow flows, the trumpet-like
co-rotating inner core, disappears completely. This change is
qualitative in nature, with a transition from concave to convex
shapes of the iso-velocity lines. In addition we note an increase
of slip near the driving disk --- while for slow flows, the
normalized angular velocity $\omega_S$ reaches 1 near the disk,
for the fastest flow $\omega_S$ has a maximum of 0.7.

The predicted flow field $\omega_D(r,z)$ for slow dry flows with
$H/R_s \approx 0.5$ is shown in
Fig.~\ref{fig:submstruct:concoverview}a -- see equations 1,2,6 and
7 from Ref~\cite{2010_softmatter_dijksman}. The similarity to the
slowest flow profile, Fig.~\ref{fig:submstruct:concoverview}b,
$\Omega = 8.3\times10^{-5}$ rps, is striking, and is confirmed in
a scatter plot of $\omega_S(r,z)$ vs $\omega_D(r,z)$, where all
data for $\Omega = 8.3\times10^{-5}$ rps (square) collapses on a
straight line
--- see Fig.~\ref{fig:submstruct:concoverview}g.  We conclude that
the flow profiles of slowly sheared gravitational suspension and
dry granular media are indistinguishable.

The predicted flow field $\omega_N(r,z)$ for Newtonian flows with
$H/R_s \approx 0.5$ is determined by a finite element software
package (COMSOL) to solve the steady state Navier-Stokes equations
for an incompressible Newtonian fluid, and is shown in
Fig.~\ref{fig:submstruct:concoverview}f. The similarity between
the  measured suspension flows for large $\Omega$ and the
Newtonian flow is, again, striking, and is confirmed in a scatter
plot of $\omega_S(r,z)$ vs $\omega_N(r,z)$, where all data for
$\Omega = 8.3\times10^{-2}$ rps (circles) tends to a straight line
(Fig.~\ref{fig:submstruct:concoverview}h).

The crossover from frictional to viscous behavior can be
quantified further by calculating as function of $\Omega$ the
total mean squared deviation ($\chi^2$) obtained from a linear fit
of the measured flow profiles $\omega_S(r,z)$ to the predicted dry
($\omega_D(r,z)$) and viscous ($\omega_N(r,z)$) flows, as shown in
Fig.~\ref{fig:submstruct:concoverview}i. We conclude that the flow
profiles of gravitational suspensions show a crossover from
frictional, granular behavior to viscous flow upon increasing the
driving rate.


\begin{figure}[t!]
    \begin{center}
        \includegraphics[width=8.5cm]{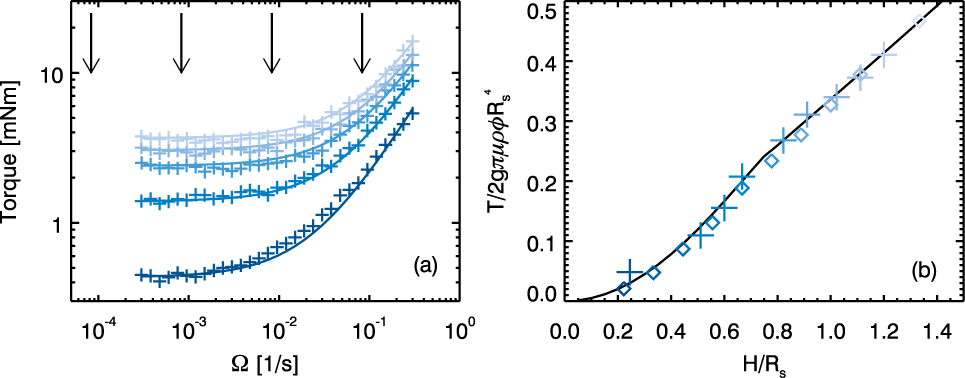}
        \caption{(Color online) (a) Rheology of 4.6 mm acrylic particles in pure Triton X-100.
        $T(\Omega)$ for a $H/R_s = $ 0.24, 0.51, 0.60, 0.67, 0.82, 0.91,
        1.0, 1.1, 1.2; color (intensity) indicates
        $H/R_s$. $\Omega = 3.0\times 10^{-4}$ to $0.3$. The curve is a fit
        of the form $T=T_0+C\Omega$ (see text).
        The four arrows indicate the driving rates where flow
        profiles were measured. (b) $T_0(H)$, the plateau values as a function
        $H$ compared to the prediction from Eq.~\ref{eq:intro:minim}
        with $\mu=0.57 \pm 0.03$ for the dry ($\diamond$) and $\mu=0.59 \pm 0.03$ for the suspension (+)
        case.
        \label{fig:submrheo:thsuspension}}
    \end{center}
\end{figure}

{\em Rheology ---} Can we find the same crossover between these
two regimes in the rheology? We measure the average driving torque
as a function of filling height $H$ and driving rate $\Omega$, in
order to connect the rheology to the findings for the flow
profiles discussed above~\cite{note_01}. Since index matching is
not necessary, we use pure Triton X-100 as interstitial fluid; we
use the same particles as before and keep the temperature fixed at
25$^{\circ}$. In each experiment, $\Omega$ is incremented from low
to high values; each data point is obtained by averaging over
three or more rotations (transients occur over much smaller
strains).

In Fig.~\ref{fig:submrheo:thsuspension}a we show $T(H,\Omega)$ for
several different suspension filling heights. We conclude that the
trends in the rheology are similar for all filling heights. First,
we observe a rate independent regime at small $\Omega$, which
corresponds to the range where we observed flow profiles similar
to the dry case. Moreover, the overall stress depends on filling
height, which we will show below to be consistent with a pressure
dependence. Second, the stresses become rate dependent for $\Omega
\sim 0.01$ rps, and for larger rotation rates, the torque
increases linearly with $\Omega$; over the whole range of driving
rates explored, the rheological data can be well fitted as
$T=T_0+C\Omega$, consistent with the behavior predicted by
Eq.~(\ref{eq:tau}) \cite{note_01}. We note here that a comparison
of the measured torque for pure Triton and for the suspension
yields that the effective viscosity of the suspension is only
three to five times larger than  $\eta_f$. This is far below than
what would naively be expected from textbook formulae, e.g.,
Krieger-Dougherty. We have no explanation for this, but note that
in the nontrivial split bottom geometry, the suspension packing
fraction varies throughout the
material~\cite{2010_softmatter_dijksman}, which complicates the
analysis.

We will now show that the height dependent torque for slow flows,
$T_0(H)$, is well described by a prediction originally developed
for slow dry flows (Fig.~\ref{fig:submrheo:thsuspension}b). From
Eq.\ref{eq:tau} it follows that the rheology should be determined
by the local hydrostatic pressure and an effective friction
coefficient $\mu_0$. Unger and coworkers~\cite{2004_prl_unger}
used these ingredients to predict $r(z)$, the center of the shear
band of the dry split-bottom flow profiles, but their model also
gives a prediction for $T(H)$:
\begin{equation}
        T(H) = 2g\pi\rho\phi\mu_0\int_0^H (H-z)r^2 \sqrt{1+(dr/dz)^2}dz.\label{eq:intro:minim}
\end{equation}
\noindent Here $\rho$ is the density of the particles, corrected
for buoyancy in case of submersed particles, $\phi$ is the average
packing fraction ($\sim 0.59$~\cite{2008_annurevfluid_forterre})
and $\mu_0$ is the effective friction coefficient. Minimization of
Eq.~\ref{eq:intro:minim} yields a prediction for $T(H)$ which has
not been tested previously.

As shown in Fig.~\ref{fig:submrheo:thsuspension}b this prediction
agrees very well with our measurements. The single fit parameter
in the model allows to accurately extract a friction coefficient,
which we estimate as $\mu_0 \sim 0.59 \pm 0.03$. We carried out
the same measurement of $T(H)$ on dry acrylic particles
(Fig.~\ref{fig:submrheo:thsuspension}b) and obtain a friction
coefficient of $\mu_0 = 0.57 \pm 0.03$. The two friction
coefficients are identical to within the experimental error, a
fact also observed in Ref.~\cite{2007_prl_divoux}. This is strong
evidence that in the slow driving rate limit the suspension
behaves as a dry granular material and that lubrication and other
hydrodynamic effects can be ignored. Furthermore, we can conclude
that the simple frictional model by Unger correctly captures the
overall stresses.

\begin{figure}[t!]
    \begin{center}
        \includegraphics[width=7.5cm]{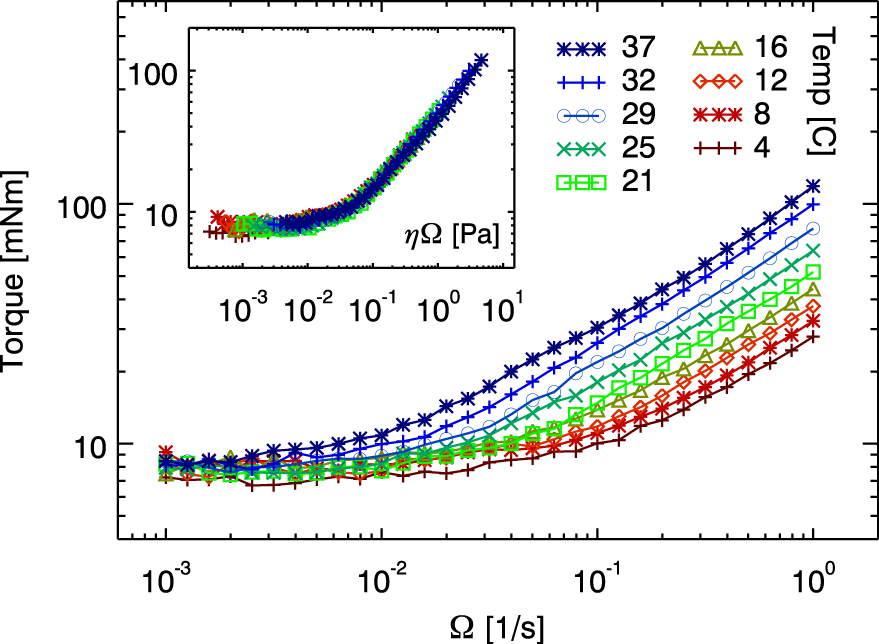}
        \caption{(Color online) The rheology of glass beads in a glycerol mixture
        at different temperatures. The inset
        shows the same data, with the abscissa rescaled with the viscosity of the
        glycerol at the given temperatures. \label{fig:submrheo:rescalerheo}}
    \end{center}
\end{figure}

We have also tested the scaling in the viscous regime, by
measuring the rheology of glass beads ($\rho$ = 2.5 $\times 10^3$
kg/m$^3$) immersed to $H/R_S = 0.4$ in glycerol for temperatures
between 4 and 37 $^{\circ}C$. The viscosity of the glycerol
mixture varies more than a decade over this temperature range, and
hence should change the rotation rate at which the viscous regime
sets in. The results are shown in
Fig.~\ref{fig:submrheo:rescalerheo}. Eq.~(\ref{eq:tau}) requires
that the data can be rescaled with the viscosity of the liquid
$\eta_f$ --- this is indeed observed in the inset of
Fig.~\ref{fig:submrheo:rescalerheo}. Note that the growth of
torque with strain rate over the larger range probed here is
somewhat slower than the simple linear prediction \cite{note_01}.

{\em Conclusions ---}  Our main finding is that with increasing
shear rate, a gravitational suspension crosses over from flowing
like a dry granular material to flowing like a viscous liquid,
consistent with recent modelling of suspensions based on the
inertial number approach. We observe this both in the full three
dimensional flow profile, which we revealed using an index matched
scanning technique, and in rheological measurements. Most of our
data can be understood based on simple scaling arguments (to
obtain the ``transition'' shear rate) or elegant minimization
principles (to obtain $\mu$ from $T(H)$). Our measurements
indicate that the shape and width of the shear band in slow
suspensions are the same as for slow dry granular flows. Whatever
the physics beyond friction necessary to produce these flow
profiles, our data shows that it is equally present in both dry
granular and gravitational suspension flows. Still, a simple
physical argument for the most prominent feature of split-bottom
shear flows --- the large width and error function shape of the
shear zone
\cite{2003_nature_fenistein,2004_prl_fenistein,2006_prl_fenistein,2006_prl_cheng,2007_epl_depken}
--- remains elusive.

Acknowledgement:  This work was supported by NSF-CTS0625890 and
NSF-DMR0907146. JAD, EW and MvH acknowledge funding from the Dutch
physics foundation FOM. We thank Krisztian Ronaszegi for the
design and construction of the 3D imaging system of the
split-bottom shear cell, and Jeroen Mesman for outstanding
technical assistance in the construction of the rheological setup.



\begin{thebibliography}{99}
\expandafter\ifx\csname
natexlab\endcsname\relax\def\natexlab#1{#1}\fi
\expandafter\ifx\csname bibnamefont\endcsname\relax
  \def\bibnamefont#1{#1}\fi
\expandafter\ifx\csname bibfnamefont\endcsname\relax
  \def\bibfnamefont#1{#1}\fi
\expandafter\ifx\csname citenamefont\endcsname\relax
  \def\citenamefont#1{#1}\fi
\expandafter\ifx\csname url\endcsname\relax
  \def\url#1{\texttt{#1}}\fi
\expandafter\ifx\csname
urlprefix\endcsname\relax\def\urlprefix{URL }\fi
\providecommand{\bibinfo}[2]{#2}
\providecommand{\eprint}[2][]{\url{#2}}

\bibitem[{\citenamefont{Coussot et~al.}(1999)\citenamefont{Ancey, Cristophe, and
  Coussot, Philippe}}]{1999_cracadparis_ancey}
\bibinfo{author}{\bibfnamefont{C.}~\bibnamefont{Ancey}} \bibnamefont{and}
  \bibinfo{author}{\bibfnamefont{P.} \bibnamefont{Coussot}},
  \bibinfo{journal}{C.R. Acad. Paris} \textbf{\bibinfo{volume}{327}},
  \bibinfo{pages}{515} (\bibinfo{year}{1999}).

\bibitem[{\citenamefont{Cassar et~al.}(2005)\citenamefont{Cassar, Nicolas, and
  Pouliquen}}]{2005_physfluids_cassar}
\bibinfo{author}{\bibfnamefont{C.}~\bibnamefont{Cassar}},
  \bibinfo{author}{\bibfnamefont{M.}~\bibnamefont{Nicolas}}, \bibnamefont{and}
  \bibinfo{author}{\bibfnamefont{O.}~\bibnamefont{Pouliquen}},
  \bibinfo{journal}{Phys. Fluids} \textbf{\bibinfo{volume}{17}},
  \bibinfo{eid}{103301} (\bibinfo{year}{2005}).

\bibitem[{\citenamefont{Siavoshi et~al.}(2006)\citenamefont{Siavoshi, Orpe, and
  Kudrolli}}]{2006_pre_siavoshi}
\bibinfo{author}{\bibfnamefont{S.}~\bibnamefont{Siavoshi}},
  \bibinfo{author}{\bibfnamefont{A.~V.} \bibnamefont{Orpe}}, \bibnamefont{and}
  \bibinfo{author}{\bibfnamefont{A.}~\bibnamefont{Kudrolli}},
  \bibinfo{journal}{Phys. Rev. E} \textbf{\bibinfo{volume}{73}},
  \bibinfo{pages}{010301} (\bibinfo{year}{2006}).

  \bibitem[{\citenamefont{Lemaitre et~al.}(2009)\citenamefont{Lemaitre, Roux, and
  Chevoir}}]{2009_rheolacta_lemaitre}
\bibinfo{author}{\bibfnamefont{A.}~\bibnamefont{Lemaitre}},
  \bibinfo{author}{\bibfnamefont{J.-N.} \bibnamefont{Roux}}, \bibnamefont{and}
  \bibinfo{author}{\bibfnamefont{F.}~\bibnamefont{Chevoir}},
  \bibinfo{journal}{Rheol. Acta} \textbf{\bibinfo{volume}{48}},
  \bibinfo{pages}{925} (\bibinfo{year}{2009}).

\bibitem[{\citenamefont{Pailha and Pouliquen}(2009)}]{2009_jfm_pailha}
\bibinfo{author}{\bibfnamefont{M.}~\bibnamefont{Pailha}} \bibnamefont{and}
  \bibinfo{author}{\bibfnamefont{O.}~\bibnamefont{Pouliquen}},
  \bibinfo{journal}{J. Fluid Mech.} \textbf{\bibinfo{volume}{633}},
  \bibinfo{pages}{115} (\bibinfo{year}{2009}).

\bibitem[{\citenamefont{Frey and Church}(2009)}]{2009_science_frey}
\bibinfo{author}{\bibfnamefont{P.}~\bibnamefont{Frey}} \bibnamefont{and}
  \bibinfo{author}{\bibfnamefont{M.}~\bibnamefont{Church}},
  \bibinfo{journal}{Science} \textbf{\bibinfo{volume}{325}},
  \bibinfo{pages}{1509} (\bibinfo{year}{2009}).

\bibitem[{\citenamefont{Jomha et~al.}(1991)\citenamefont{Jomha, Merrington,
  Woodcock, Barnes, and Lips}}]{1991_powtechn_jomha}
\bibinfo{author}{\bibfnamefont{A.}~\bibnamefont{Jomha}} \textit{et. al.},
  \bibinfo{journal}{Powder Techn.} \textbf{\bibinfo{volume}{65}},
  \bibinfo{pages}{343 } (\bibinfo{year}{1991}).

\bibitem[{\citenamefont{Stickel and
  Powell}(2005)}]{2005_annurevfluidmech_stickel}
\bibinfo{author}{\bibfnamefont{J.~J.} \bibnamefont{Stickel}} \bibnamefont{and}
  \bibinfo{author}{\bibfnamefont{R.}~\bibnamefont{Powell}},
  \bibinfo{journal}{Annu. Rev. Fluid Mech.} \textbf{\bibinfo{volume}{37}},
  \bibinfo{pages}{129} (\bibinfo{year}{2005}).

\bibitem[{\citenamefont{Fall et~al.}(2009)\citenamefont{Fall, Bertrand,
  Ovarlez, and Bonn}}]{2009_prl_fall}
\bibinfo{author}{\bibfnamefont{A.}~\bibnamefont{Fall}},
  \bibinfo{author}{\bibfnamefont{F.}~\bibnamefont{Bertrand}},
  \bibinfo{author}{\bibfnamefont{G.}~\bibnamefont{Ovarlez}}, \bibnamefont{and}
  \bibinfo{author}{\bibfnamefont{D.}~\bibnamefont{Bonn}},
  \bibinfo{journal}{Phys. Rev. Lett.} \textbf{\bibinfo{volume}{103}},
  \bibinfo{eid}{178301} (\bibinfo{year}{2009}).

\bibitem[{\citenamefont{Einstein}(1905)}]{1906_annphys_einstein}
\bibinfo{author}{\bibfnamefont{A.}~\bibnamefont{Einstein}},
  \bibinfo{journal}{Ann. Physik} \textbf{\bibinfo{volume}{19}},
  \bibinfo{pages}{289} (\bibinfo{year}{1905}).

\bibitem[{\citenamefont{Jaeger et~al.}(1996)\citenamefont{Jaeger, Nagel, and
  Behringer}}]{1996_revmodphys_jaeger}
\bibinfo{author}{\bibfnamefont{H.~M.} \bibnamefont{Jaeger}},
  \bibinfo{author}{\bibfnamefont{S.~R.} \bibnamefont{Nagel}}, \bibnamefont{and}
  \bibinfo{author}{\bibfnamefont{R.~P.} \bibnamefont{Behringer}},
  \bibinfo{journal}{Rev. Mod. Phys.} \textbf{\bibinfo{volume}{68}},
  \bibinfo{pages}{1259} (\bibinfo{year}{1996}).

\bibitem[{\citenamefont{Forterre and
  Pouliquen}(2008)}]{2008_annurevfluid_forterre}
\bibinfo{author}{\bibfnamefont{Y.}~\bibnamefont{Forterre}} \bibnamefont{and}
  \bibinfo{author}{\bibfnamefont{O.}~\bibnamefont{Pouliquen}},
  \bibinfo{journal}{Annu. Rev. Fluid Mech.} \textbf{\bibinfo{volume}{40}},
  \bibinfo{pages}{1} (\bibinfo{year}{2008}).

\bibitem[{\citenamefont{Tardos et~al.}(1998)\citenamefont{Tardos, Khan, and
  Schaeffer}}]{1998_physfluids_tardos}
\bibinfo{author}{\bibfnamefont{G.~I.} \bibnamefont{Tardos}},
  \bibinfo{author}{\bibfnamefont{M.~I.} \bibnamefont{Khan}}, \bibnamefont{and}
  \bibinfo{author}{\bibfnamefont{D.~G.} \bibnamefont{Schaeffer}},
  \bibinfo{journal}{Phys. Fluids} \textbf{\bibinfo{volume}{10}},
  \bibinfo{pages}{335} (\bibinfo{year}{1998}).

\bibitem[{\citenamefont{MiDi}(2004)}]{2004_epje_gdrmidi}
\bibinfo{author}{\bibfnamefont{G.}~\bibnamefont{MiDi}}, \bibinfo{journal}{Eur.
  Phys. J. E} \textbf{\bibinfo{volume}{14}}, \bibinfo{pages}{341}
  (\bibinfo{year}{2004}).

\bibitem[{\citenamefont{Jop et~al.}(2006)\citenamefont{Jop, Forterre, and
  Pouliquen}}]{2006_nature_jop}
\bibinfo{author}{\bibfnamefont{P.}~\bibnamefont{Jop}},
  \bibinfo{author}{\bibfnamefont{Y.}~\bibnamefont{Forterre}}, \bibnamefont{and}
  \bibinfo{author}{\bibfnamefont{O.}~\bibnamefont{Pouliquen}},
  \bibinfo{journal}{Nature} \textbf{\bibinfo{volume}{441}}, \bibinfo{eid}{727}
  (\bibinfo{year}{2006}).

\bibitem[{\citenamefont{Pailha et~al.}(2008)\citenamefont{Pailha, Nicolas, and
  Pouliquen}}]{2008_physfluid_pailha}
\bibinfo{author}{\bibfnamefont{M.}~\bibnamefont{Pailha}},
  \bibinfo{author}{\bibfnamefont{M.}~\bibnamefont{Nicolas}}, \bibnamefont{and}
  \bibinfo{author}{\bibfnamefont{O.}~\bibnamefont{Pouliquen}},
  \bibinfo{journal}{Phys. Fluids} \textbf{\bibinfo{volume}{20}},
  \bibinfo{eid}{111701} (\bibinfo{year}{2008}).

\bibitem[{\citenamefont{Divoux and G\'eminard}(2007)}]{2007_prl_divoux}
\bibinfo{author}{\bibfnamefont{T.}~\bibnamefont{Divoux}} \bibnamefont{and}
  \bibinfo{author}{\bibfnamefont{J.-C.} \bibnamefont{G\'eminard}},
  \bibinfo{journal}{Phys. Rev. Lett.} \textbf{\bibinfo{volume}{99}},
  \bibinfo{pages}{258301} (\bibinfo{year}{2007}).

\bibitem[{\citenamefont{Doppler et~al.}(2007)\citenamefont{Doppler, Gondret,
  Loiseleux, Meyer, and Rabaud}}]{2007_jfm_doppler}
  \bibinfo{author}{\bibfnamefont{D.}~\bibnamefont{Doppler}} \textit{et. al.},
  \bibinfo{journal}{J. Fluid Mech.} \textbf{\bibinfo{volume}{577}},
  \bibinfo{pages}{161} (\bibinfo{year}{2007}).

\bibitem[{\citenamefont{Fenistein and van Hecke}(2003)}]{2003_nature_fenistein}
\bibinfo{author}{\bibfnamefont{D.}~\bibnamefont{Fenistein}} \bibnamefont{and}
  \bibinfo{author}{\bibfnamefont{M.}~\bibnamefont{van Hecke}},
  \bibinfo{journal}{Nature} \textbf{\bibinfo{volume}{425}}, \bibinfo{eid}{256}
 (\bibinfo{year}{2003}).

\bibitem[{\citenamefont{Fenistein et~al.}(2004)\citenamefont{Fenistein, van~de
  Meent, and van Hecke}}]{2004_prl_fenistein}
\bibinfo{author}{\bibfnamefont{D.}~\bibnamefont{Fenistein}},
  \bibinfo{author}{\bibfnamefont{J.-W.} \bibnamefont{van~de Meent}},
  \bibnamefont{and} \bibinfo{author}{\bibfnamefont{M.}~\bibnamefont{van
  Hecke}}, \bibinfo{journal}{Phys. Rev. Lett.} \textbf{\bibinfo{volume}{92}},
  \bibinfo{eid}{094301} (\bibinfo{year}{2004}).

\bibitem[{\citenamefont{Fenistein et~al.}(2006)\citenamefont{Fenistein, van~de
  Meent, and van Hecke}}]{2006_prl_fenistein}
\bibinfo{author}{\bibfnamefont{D.}~\bibnamefont{Fenistein}},
  \bibinfo{author}{\bibfnamefont{J.-W.} \bibnamefont{van~de Meent}},
  \bibnamefont{and} \bibinfo{author}{\bibfnamefont{M.}~\bibnamefont{van
  Hecke}}, \bibinfo{journal}{Phys. Rev. Lett.} \textbf{\bibinfo{volume}{96}},
  \bibinfo{eid}{118001} (\bibinfo{year}{2006}).

\bibitem[{\citenamefont{Cheng et~al.}(2006)\citenamefont{Cheng, Lechman,
  Fernandez-Barbero, Grest, Jaeger, Karczmar, M\"{o}bius, and
  Nagel}}]{2006_prl_cheng}
\bibinfo{author}{\bibfnamefont{X.}~\bibnamefont{Cheng}} \textit{et. al.}
  \bibinfo{journal}{Phys. Rev. Lett.} \textbf{\bibinfo{volume}{96}},
  \bibinfo{eid}{038001} (\bibinfo{year}{2006}).

\bibitem[{\citenamefont{Depken et~al.}(2007)\citenamefont{Depken, Lechman, van
  Hecke, van Saarloos, and Grest}}]{2007_epl_depken}
\bibinfo{author}{\bibfnamefont{M.}~\bibnamefont{Depken}} \textit{et. al.},
  \bibinfo{journal}{EPL} \textbf{\bibinfo{volume}{78}}, \bibinfo{pages}{58001} (\bibinfo{year}{2007}).

\bibitem[{\citenamefont{Ries et~al.}(2007)\citenamefont{Ries, Wolf, and
  Unger}}]{2007_pre_ries}
\bibinfo{author}{\bibfnamefont{A.}~\bibnamefont{Ries}},
  \bibinfo{author}{\bibfnamefont{D.~E.} \bibnamefont{Wolf}}, \bibnamefont{and}
  \bibinfo{author}{\bibfnamefont{T.}~\bibnamefont{Unger}},
  \bibinfo{journal}{Phys. Rev. E} \textbf{\bibinfo{volume}{76}},
  \bibinfo{eid}{051301} (\bibinfo{year}{2007}).

\bibitem[{\citenamefont{Dijksman et~al.}(2004)\citenamefont{Dijksman, Joshua A.,
  and Hecke}}]{2010_softmatter_dijksman}
  \bibinfo{author}{\bibfnamefont{J.A.}~\bibnamefont{Dijksman}},
  \bibnamefont{and} \bibinfo{author}{\bibfnamefont{M.}~\bibnamefont{van Hecke}},
  \bibinfo{journal}{Soft Matter} \textbf{\bibinfo{volume}{6}}, \bibinfo{pages}{2901}
  (\bibinfo{year}{2010}).

\bibitem[{\citenamefont{Sakaie et~al.}(2004)\citenamefont{Sakaie, Fenistein,
Carroll, van Hecke, and Umbanhowar}}]{2008_epl_sakaie}
  \bibinfo{author}{\bibfnamefont{K.}~\bibnamefont{Sakaie}} \textit{et. al.}
  \bibinfo{journal}{EPL} \textbf{\bibinfo{volume}{84}}, \bibinfo{pages}{49902}
  (\bibinfo{year}{2008}).

\bibitem[]{note_01}
  \bibinfo{note}{Due to the variations in the local pressure, the strain rate
    and the unknown local stress field, we can only access the global
    rheological quantities, and the mapping from $\Omega$ to strain
    rate $\gammadot$ is only approximate.}


\bibitem[{\citenamefont{Krishnan et~al.}(1996)\citenamefont{Krishnan, Beimfohr,
  and Leighton}}]{1996_jfm_krishnan}
\bibinfo{author}{\bibfnamefont{G.~P.} \bibnamefont{Krishnan}},
  \bibinfo{author}{\bibfnamefont{S.}~\bibnamefont{Beimfohr}}, \bibnamefont{and}
  \bibinfo{author}{\bibfnamefont{D.~T.} \bibnamefont{Leighton}},
  \bibinfo{journal}{J. Fluid Mech.} \textbf{\bibinfo{volume}{321}},
  \bibinfo{pages}{371} (\bibinfo{year}{1996}).

\bibitem[{\citenamefont{Tsai et~al.}(2003)\citenamefont{Tsai, Voth, and
  Gollub}}]{2003_prl_tsai}
\bibinfo{author}{\bibfnamefont{J.-C.} \bibnamefont{Tsai}},
  \bibinfo{author}{\bibfnamefont{G.~A.} \bibnamefont{Voth}}, \bibnamefont{and}
  \bibinfo{author}{\bibfnamefont{J.~P.} \bibnamefont{Gollub}},
  \bibinfo{journal}{Phys. Rev. Lett.} \textbf{\bibinfo{volume}{91}},
  \bibinfo{pages}{064301} (\bibinfo{year}{2003}).

\bibitem[{\citenamefont{Slotterback et~al.}(2008)\citenamefont{Slotterback,
  Toiya, Goff, Douglas, and Losert}}]{2008_prl_slotterback}
\bibinfo{author}{\bibfnamefont{S.}~\bibnamefont{Slotterback}} \textit{et.
al.},
  \bibinfo{journal}{Phys. Rev. Lett.} \textbf{\bibinfo{volume}{101}},
  \bibinfo{eid}{258001} (\bibinfo{year}{2008}).

\bibitem[{\citenamefont{Dijksman et~al.}()\citenamefont{Dijksman, Lorincz,
  Rietz, van Hecke, and Losert}}]{2011_revsciinstr_dijksman}
\bibinfo{author}{\bibfnamefont{J.}~\bibnamefont{Dijksman}} \textit{et. al.},
  \bibinfo{journal}{in preparation}.

\bibitem[{\citenamefont{Koval et~al.}(2009)\citenamefont{Koval, Roux, Corfdir,
  and Chevoir}}]{2009_pre_koval}
\bibinfo{author}{\bibfnamefont{G.}~\bibnamefont{Koval}},
  \bibinfo{author}{\bibfnamefont{J.-N.} \bibnamefont{Roux}},
  \bibinfo{author}{\bibfnamefont{A.}~\bibnamefont{Corfdir}}, \bibnamefont{and}
  \bibinfo{author}{\bibfnamefont{F.}~\bibnamefont{Chevoir}},
  \bibinfo{journal}{Phys. Rev. E} \textbf{\bibinfo{volume}{79}},
  \bibinfo{eid}{021306} (\bibinfo{year}{2009}).

\bibitem[{\citenamefont{Unger et~al.}(2004)\citenamefont{Unger, T\"or\"ok,
  Kert\'esz, and Wolf}}]{2004_prl_unger}
\bibinfo{author}{\bibfnamefont{T.}~\bibnamefont{Unger}},
  \bibinfo{author}{\bibfnamefont{J.}~\bibnamefont{T\"or\"ok}},
  \bibinfo{author}{\bibfnamefont{J.}~\bibnamefont{Kert\'esz}},
  \bibnamefont{and} \bibinfo{author}{\bibfnamefont{D.~E.} \bibnamefont{Wolf}},
  \bibinfo{journal}{Phys. Rev. Lett.} \textbf{\bibinfo{volume}{92}},
  \bibinfo{pages}{214301} (\bibinfo{year}{2004}).



\end{thebibliography}

\end{document}